\documentclass[11pt,preprint]{aastex}
\usepackage{amsmath}
\usepackage{amssymb}
\usepackage{epsfig}

\newcommand{\cms}{\,{\rm cm$^{-2}$}\,}
\newcommand{\cmc}{\,{\rm cm$^{-3}$}\,}
\newcommand{\kms}{\,{\rm km\,s$^{-1}$}\,}
\newcommand{\kmsmpc}{\,{\rm km\,s$^{-1}$\,Mpc$^{-1}$}\,}

\newcommand{\etal}{{ et~al.~}}
\newcommand{\expunit}{\,{\rm photons\,s$^{-1}$\,cm$^{-2}$\,arcsec$^{-2}$}\,}

\newcommand{\Zs}{Z_\odot}

\shorttitle{Infall of NGC 1404 into Fornax}
\begin{document}


\title{Infall of the Elliptical Galaxy NGC~1404 into the Fornax Cluster}

\author{M. Machacek, A. Dosaj, W. Forman, C. Jones \\ 
  M. Markevitch,  A. Vikhlinin, A. Warmflash, R. Kraft}
\affil{Harvard-Smithsonian Center for Astrophysics \\ 
       60 Garden Street, Cambridge, MA 02138 USA
\email{mmachacek@cfa.harvard.edu}}

\begin{abstract}
We use three Chandra observations, totalling $134.3$\,ks,
to constrain the dynamical motion 
of NGC~1404 falling towards the dominant elliptical NGC~1399 
through the Fornax cluster gas. 
The surface brightness profile of NGC~1404 shows a sharp edge 
at $\sim 8$\,kpc from its center in 
the direction of NGC~1399, characteristic of jump-like temperature and
density discontinuities from ram-pressure stripping of the galaxy gas 
caused by its motion through the surrounding ICM. 
We find the temperature of the galaxy gas inside the edge is 
$\sim 2.8$ times cooler
($kT =0.55^{+0.01}_{-0.02}$\,keV with abundance 
$A=0.73^{+0.65}_{-0.16}\Zs$) than the 
cluster gas ($kT = 1.53^{+0.10}_{-0.13}$\,keV, 
$A=0.42^{+0.2}_{-0.13}\Zs$). We use the shape of the surface
brightness profile across the edge to fit the position of the edge, 
the power law behavior of NGC~1404's density distribution in the leading 
direction, and the density discontinuity at the edge. 
The electron  
density inside the edge ($3.9 - 4.3 \times 10^{-3}$\cmc) depends strongly
on the gas abundance; while the density of the ICM 
($7 - 8 \times 10^{-4}$\cmc) depends strongly on the assumed 
geometry (relative distance) between NGC~1404 and NGC~1399.
The corresponding pressure jump of $1.7-2.1$ across the leading edge
of the galaxy and the cluster free-stream region implies near sonic
motion (Mach number $0.83 - 1.03$) for NGC~1404 
with velocity $531-657$\kms relative
to the surrounding cluster gas. The inclination angle of the motion 
inferred using the relative radial velocity  
between NGC~1404 and NGC~1399 as representative of that between NGC~1404 and
the cluster ICM, is uncomfortably large ($\gtrsim 40^\circ$)
given the sharpness of the
surface brightness edge, suggesting either a non-zero impact parameter
between NGC~1404 and NGC~1399 or that NGC~1399 is also moving radially
with respect to the cluster ICM.

\end{abstract}



\section{Introduction}
\label{sec:introduction}

Our early picture of galaxy clusters, as simple, dynamically relaxed 
systems, has been changed dramatically over the past decade by 
X-ray images taken with the Einstein Observatory, 
ROSAT, BBXRT, and ASCA. These images show that
galaxy clusters are complex systems with extensive substructure 
 and mergers on time sales of $\sim 10^9$ years. 
(see e.g. Forman \etal 2001, 2002 and references therein). The excellent 
angular resolution of the Chandra X-ray Observatory made possible the 
detailed study of these substructures, particularly near the 
centers of clusters, revealing surprising new signatures for 
their interactions.
Notably ram pressure stripping of subcluster gas due to its motion
through the intracluster medium (ICM) was found to produce prominant 
X-ray surface brightness discontinuities from cold fronts (dense
surviving subcluster cores), rather than from the shocked gas also expected
from such motion (Forman \etal 2001, 2002; Vikhlinin \etal 2001; 
Markevitch \etal 2000; Heinz \etal 2003). Furthermore, 
high quality X-ray measurements of the geometry and gas
properties along the interface between the cold front and the ICM
were used to determine the dynamical motion of the cold front 
 through the cluster (Vikhlinin \etal 2001; Mazzotta \etal 2001).
 
Galaxies in these dynamically rich cluster environments are 
subject to both tidal and hydrodynamical interactions that 
significantly affect their evolution (see, 
Gnedin  2003; Acreman \etal 2003 and references therein). 
The Chandra X-ray Observatory has also made possible
detailed studies of hot gas in member galaxies of
nearby clusters in interaction with each other and their gaseous
environment. Just as in the case of  
larger cluster substructure, the action of ram pressure acting 
on gas within an individual galaxy due to the galaxy's motion 
through the ICM produces  sharp X-ray surface brightness 
discontinuities along the galaxy's leading edge associated with 
cold fronts, as well as trailing wakes and comet-like debris tails of 
gas swept from the infalling system (e.g. see Forman \etal 1979, 
White \etal 1991, Rangarajan \etal 1995 for M86, and Irwin \&
Sarazin 1996 for NGC~4472, both in Virgo; and Wang \etal 2004 
for C153 in Abell 2125). 
Thus, in addition 
to elucidating the  hydrodynamics of ram pressure 
stripping itself, high quality X-ray measurements of the geometry of
these features combined with measurements of the gas
temperatures, abundances, and densities along the interface between
the cold front and the ICM may be used to constrain the three dimensional
(transverse as well as radial) motion of the
galaxy through the cluster (Dosaj \etal 2002). 

\begin{figure}[t]
\plotone{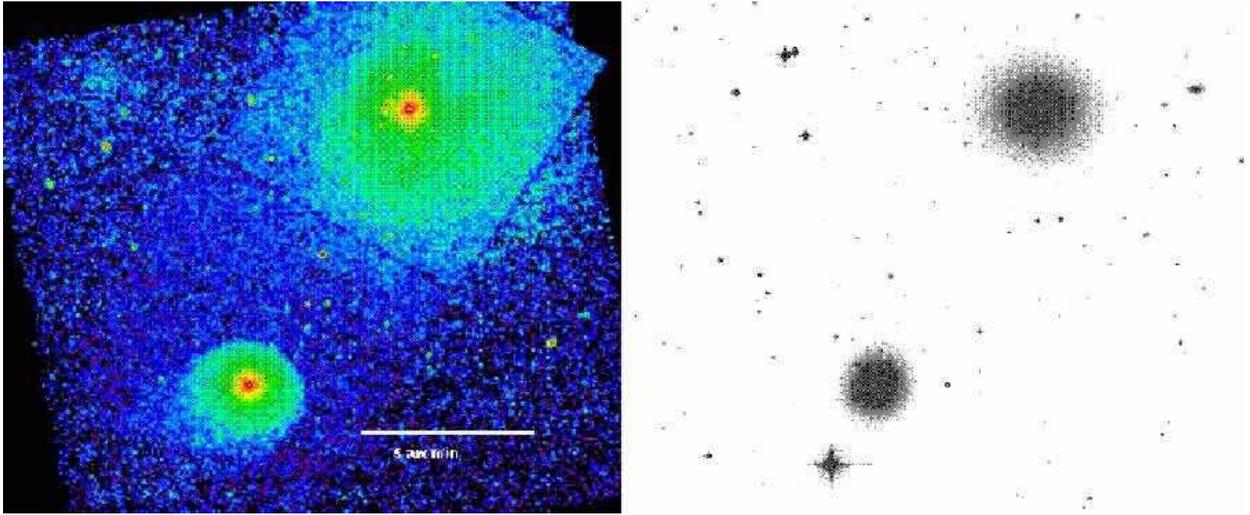}
\caption{The $0.3-2.0$\,keV ACIS-S X-ray (left) and DSS (right)
images of NGC1404 (SE, lower left) and NGC1399 (NW, upper right)
galaxies in the Fornax cluster. The X-ray images have been binned by 
$2$ ($0''.984$ pixels), background subtracted, exposure corrected and 
coadded in WCS coordinates before smoothing with a $2''$ Gaussian.
The image color scale has logarithmic stretch from 
$ 10^{-9}$\expunit to $6 \times 10^{-6}$\expunit. The distance scale
bar is $5$\,arcmin.
The bright elliptical galaxy NGC~1404 exhibits a surface
brightness discontinuity along its northwestern edge 
at a radius of $\sim 1'.4$($8$\,kpc) from its center.}
\label{fig:fornax}
\end{figure}

In this paper we present this analysis for one such galaxy,
NGC~1404 (shown in Figure \ref{fig:fornax}), 
that is undergoing ram pressure stripping as it falls inwards 
towards the Fornax cluster center in the direction of the dominant 
elliptical galaxy NGC~1399. 
Both NGC~1399 and NGC~1404 are well studied in X-rays. 
Early observations of NGC~1399 using 
EXOSAT, Einstein IPC, Ginga, and BBXRT 
(see Jones \etal 1997 and references therein)
established the presence and mean temperature of its X-ray emitting
halo. More recent measurements by ASCA, ROSAT, XMM-Newton and Chandra
mapped the temperature and density profiles for the galaxy and the 
surrounding Fornax ICM (Jones \etal 1997; Buote 2002; Paolillo \etal 2002;
O'Sullivan \etal 2003; Scharf \etal 2004). 
NGC~1404 has been observed with ASCA (Loewenstein \etal 1994;
 Buote \& Fabian 1998), ROSAT PSPC 
(Jones \etal 1997; Paolillo \etal 2002; O'Sullivan \etal 2003)
and Chandra (Forman \etal 2002; Dosaj \etal 2002; Scharf \etal 2004; 
this work). The temperature of diffuse gas in NGC~1404 was found to be
approximately isothermal with $kT \sim 0.6$\,keV. The higher
temperature ($kT \sim 0.75$\,keV) found by 
Loewenstein \etal (1994) was  
most likely a result of the large radius ($3'$) aperture used, due to
the ASCA point spread function, that
extended well beyond the effective radius of NGC~1404. With the much 
better ROSAT PSPC resolution, Jones \etal (1997) found an asymmetric
surface brightness distribution strongly suggesting that NGC~1404 
was undergoing ram pressure stripping. 
Also using ROSAT PSPC data, Paolillo \etal (2002) measured 
the surface brightness asymmetry between the leading (northwest) sector and 
trailing (southeast) sector of NGC~1404, showing the steep gradient in
the surface brightness in the northwest sector at  $r \sim 90''$~($8$\,kpc);
while O'Sullivan \etal (2003) 
observed a temperature jump for $r > 2'$, across the
leading edge of the galaxy and the surrounding Fornax ICM.

In Figure \ref{fig:fornax} we present the $0.3-2$\,keV coadded image from 
three Chandra observations, totalling $134.3$\,ks,  
(detailed in Section \ref{sec:obs}) of NGC~1404 
falling into the dominant cluster elliptical NGC~1399 (left
panel). For comparison we also show the Digitized Sky Survey optical image
matched in WCS coordinates (right panel). NGC~1399 is located in the 
northwest (upper right) corner of each panel; while the bright
elliptical NGC~1404 is located  
$\sim 54$\,kpc ($9'.8$) in the plane of the sky to the  
southeast (lower left corner). The X-ray images have been individually
background subtracted and corrected for telescope vignetting and
spatial efficiency variations by means of
exposure maps generated with standard CIAO tools assuming a fixed 
spectral energy of $0.9$\,keV. 
We clearly see the sharp edge in the surface brightness on the northwest 
side of NGC~1404 (in the direction of NGC~1399), formed by gas bound to
NGC~1404 shaped by the ram pressure of the cluster gas, and
also the characteristic tail of ram-pressure-stripped gas to
the galaxy's southeast. 
We use data from these three Chandra
observations to measure the temperature and density of gas on either 
side of the surface brightness edge and calculate the dynamical motion 
of NGC~1404 relative to the Fornax ICM. This is an extension of
earlier work on the Fornax cluster by Dosaj \etal (2002) using the 
Chandra ACIS-S observation (Obsid 319) to estimate the motion of 
NGC~1404 towards NGC~1399, and complements the recent, similar  
work by Scharf \etal (2004) using ACIS-I observations.  Both
assumed that the motion was primarily in the plane of the sky and, 
from the jump in gas pressure, obtained infall velocities of 
$466 \pm 38$\kms (Dosaj \etal 2002) and $660 \pm 260$\kms (Scharf
\etal 2004). 

Our discussion is organized as follows:  
In Section \ref{sec:obs} we describe the observations and our 
data reduction and processing procedures.
In Section \ref{sec:results} we discuss our analysis method and 
main results. In \ref{sec:temps} we use spectral analyses to 
determine the temperature inside the surface brightness edge 
 for NGC~1404 and outside the edge in the surrounding 
Fornax ICM showing that gas within the surface brightness edge is 
a factor $\sim 3$ cooler than the surrounding cluster gas.
In Section \ref{sec:edges} we use coadded Chandra images from the 
three observations of NGC~1404  to extract the X-ray surface
brightness profile across the edge and model the geometry of the 
cold front. We combine these data with the spectral information to 
construct the corresponding electron density profile across the edge, 
and in Section \ref{sec:velocity} we calculate the  velocity of 
NGC~1404 relative to the Fornax ICM. In   
Section \ref{sec:conclude} we briefly summarize our results.
Unless otherwise indicated all errors are $90\%$ confidence levels and 
coordinates are J2000.
Assuming a Hubble constant $H_0=75$\kmsmpc in the currently preferred
flat $\Lambda$CDM cosmology ($\Omega_m=0.3$, $\Omega_\Lambda = 0.7$)
and a redshift $z=0.004753$, the redshift of the dominant elliptical 
galaxy NGC~1399, as representative for the Fornax cluster, the
luminosity distance to the cluster is $19$\,Mpc (Paolillo \etal 2002)
and $1'$ corresponds to a distance scale of $5.49$\,kpc. 


\section{Observations and Data Reduction}
\label{sec:obs}

Our data consist of three observations of galaxies NGC~1404 and
NGC~1399 in Fornax taken with the Chandra X-ray
Observatory using the Advanced CCD Imaging Spectrometer array 
(ACIS, Garmire \etal 1992; Bautz \etal 1998): a $57.4$~ksec exposure
taken on 18-19 January 2000 (when the focal plane
temperature of the instrument was $-110^\circ$\,C) using chips
S2  and S3 (Obsid 319), a $29.6$\,ks exposure (of NGC~1404 alone) taken on 
13 February 2003 using the S3 chip (Obsid 2942), and a $47.3$\,ks exposure 
taken on 28-29 May 2003 using the ACIS-I array (Obsid 4174). 
For Obsid 2942 and Obsid 4174 the focal plane temperature was 
$-120^\circ$\,C throughout the observation. 
Each CCD chip is a $1024 \times 1024$ pixel array 
where each pixel subtends $0''.492 \times 0''.492$ on the sky. 

The data were analyzed using the standard X-ray processing packages, 
CIAO $3.0.2$, FTOOLS and XSPEC $11.2$. 
We corrected for the charge transfer inefficiency on the 
frontside-illuminated chips caused by exposure to low energy protons
during a passage of the telescope through the Earth's radiation belts 
shortly after launch (Prigozhin \etal 2000), as well as for the
time-dependent declining efficiency of the ACIS detector due to the 
buildup of contaminants on the optical filter (Plucinsky \etal 2003)
that is important at energies below $1.5$\,keV.
Filtering removed events with bad grades 
($1$, $5$, and $7$) and any events falling on hot pixels.
Periods of anomalously high background (flares) were identified and
were removed from the data along with
periods of anomalously low count rates at the beginning and end of each
observation. This resulted in useful exposure times 
of $56,241$\,s, $29,299$\,s, and $41,478$\,s, respectively, for the 
observations. 

Backgrounds for these observations were created from the 
source free ( blank sky) background sets appropriate for the CCDs,
dates of observation and instrument configurations of interest 
(Markevitch \etal 2001)\footnote{see
http://cxc.harvard.edu/contrib/maxim/acisbg }. 
These were the merged $201,452$\,s 
($114,886$\,s) Period B source free dataset for the S2(S3) chip 
with ACIS-S at the aimpoint for Obsid 319,
the  merged $450,000$\,s Period D source free dataset for the S3 chip
with ACIS-S at aimpoint for Obsid 2942, and the 
merged $550,000$\,s Period D source free data set for the ACIS-I array
with ACIS-I at the aimpoint for Obsid 4174. 
Identical cleaning, energy and spatial
filters were applied to source and background data throughout, so that
the normalization of the background is set, in each case, by the ratio
of exposure times. We checked this normalization by comparing count
rates between the source and background files in the 
$9-12$\,keV range where particle background dominates. 
The source count rates were found to differ by $\lesssim 3.5\%$ from 
those of the blank sky sets for Obsid 319 and 2942, but were 
anomalously high ($\sim 16\%$) for Obsid 4174. 
We normalized the blank sky sets for that observation to agree with
the observed $9-12$\,keV source count rates. 
Thus we adopt $3.5\%$ as the remaining 
uncertainty in the background levels for the three data sets. 
We identified point sources in the $16'.8 \times 16'.8$ field of the 
combined data in the $0.3-10$\,keV energy band using a multiscale 
wavelet decomposition algorithm set with detection
threshold at $5\sigma$.
The resulting $220$ source identifications
were excluded from the following spectral and surface brightness
analyses. We refer the reader to Angelini \etal (2001) for a detailed 
analysis of the point source population of this system.

\section{Results}
\label{sec:results}

\begin{figure}[t]
\plotone{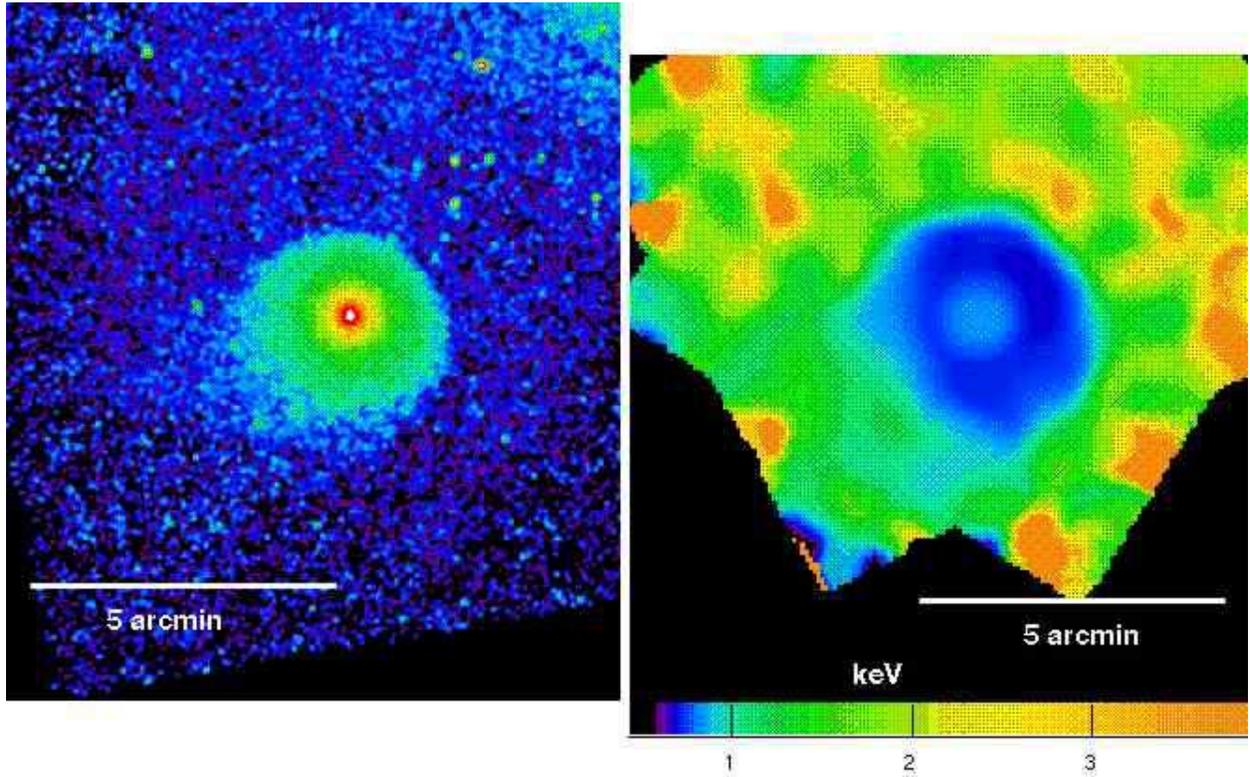}
\caption{Surface brightness (left) and temperature map (right), 
matched in WCS coordinates, for NGC~1404 and the surrounding region. 
}
\label{fig:tmap}
\end{figure}
\begin{figure}[t]
\epsscale{0.7}
\plotone{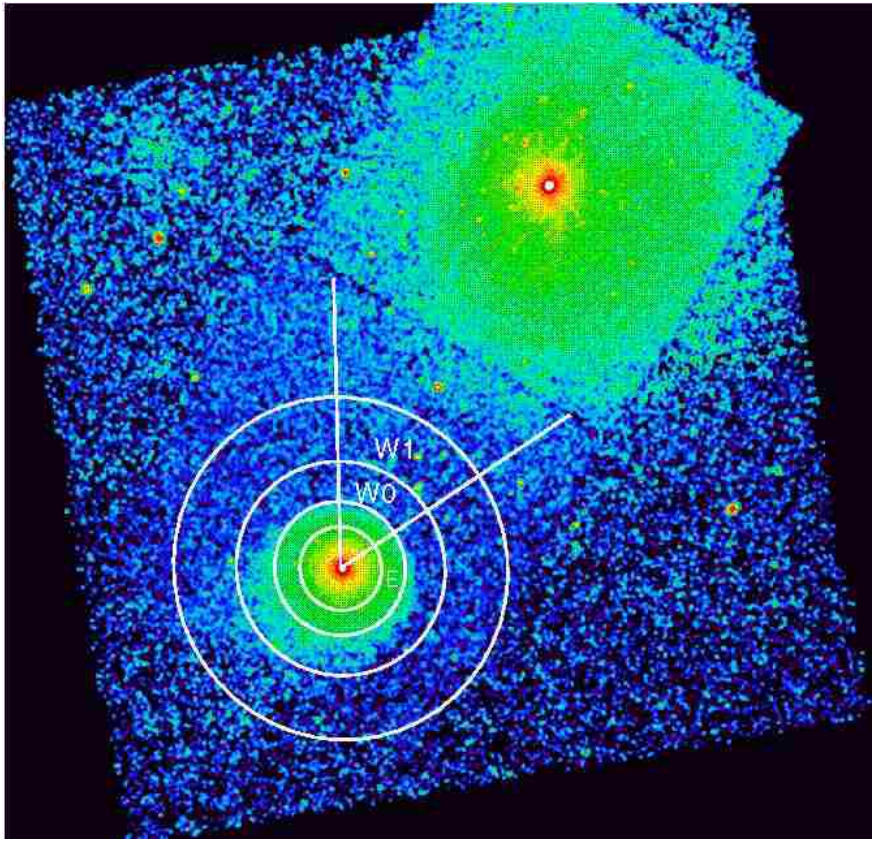}
\caption{Regions used in the edge analysis. The surface brightness 
profiles are extracted in elliptical annulli concentric to those shown
and lieing within the angular sector shown (from $35^\circ$ to $91.4^\circ$ 
measured counterclockwise from East). Spectral extraction regions 
W$0$ and W$1$ for the cluster gas outside the edge are also
constrained to lie within that angular sector; while the full
elliptical annulus E is used for the spectrum of NGC~1404 just inside 
the edge.}  
\label{fig:wedgegeom}
\end{figure}

In Figure \ref{fig:tmap} we present surface brightness and temperature 
maps of the galaxy that clearly show the 
general structure of the X-ray emission from NGC~1404  and the 
surrounding ICM, i.e a sharp, bow-like shaped drop in surface
brightness that is correlated to a similarly-shaped jump in temperature.
In the northwest direction (the leading edge) the 
bow-shaped edge is nearly circular; while at larger angles the
emission merges into a comet-like tail. In order to isolate the
leading edge, we trace the surface brightness discontinuity with an 
elliptical segment centered at WCS coordinates ($3^h38^m52.1^s$, $-35^\circ
35'39.6''$) with major(minor) axes of $8.3$($8.0$)\,kpc and position
angle  $19.94^\circ$. The center of this edge-bounding ellipse is 
offset by $\lesssim 2''.4$ to the east of NGC~1404's optical center.
We construct the surface brightness
profile for the exposure-corrected, coadded image using
elliptical annulli concentric to the edge-bounding ellipse and constrained to 
lie within an angular sector extending from $35^\circ$ to $91.4^\circ$ 
measured counterclockwise from east (See Figure \ref{fig:wedgegeom}).
For the surface brightness profile, radial intervals are chosen to 
increase(decrease) with logarithmic step size $1.05$ as the radial 
distance is increased (decreased) from the position of the bounding
ellipse. We need to obtain corresponding temperatures for gas in these
regions. However, in order to obtain sufficient statistics for
spectral analysis, we increase the logarithmic step size outside the 
bounding ellipse to $1.16$. These spectral extraction regions are 
shown in Figure \ref{fig:wedgegeom}. 
We interpret region W1 as representative of the 
undisturbed, free streaming Fornax cluster gas and region W0 as cluster gas 
directly outside the edge (Vikhlinin \etal 2001). In order to determine the 
spectral properties of gas inside the edge we use the elliptical 
annulus denoted by E in Figure \ref{fig:wedgegeom} to maximize 
statistics in order to sharpen our determination of the metal abundance 
in NGC~1404 as well as the gas temperature inside the edge. 

\subsection{Gas Temperatures}
\label{sec:temps}

Spectra from both Obsid 2942 and Obsid 4174 were fit simultaneously with
XSPEC $11.2$ over the $0.3 - 5$\,keV energy
range using APEC emission models for collisionally-ionized diffuse gas
(Smith \etal 2001) corrected for absorption using Wisconsin 
photo-electric cross-sections 
(Morrison \& McCammon 1983). Obsid 319 was not used in the spectral analysis
due to calibration uncertainties for Period B observations. 
Counts were grouped with a pre-defined grouping resulting in channels
of approximately constant logarithmic width. 
The hydrogen column density was initially fixed at its
Galactic value $1.45 \times 10^{20}$\cms (see 
http://heasarc.gsfc.nasa.gov/, Archives \& Software, nH:Column
Density) with the gas temperature and
abundance taken as free parameters. The hydrogen absorbing column was then 
allowed to vary to check the stability of the fit. The data showed no 
need for increased absorption either in the cluster gas or in the 
outer elliptical region E in NGC~1404 just inside the edge.
\begin{figure}[t]
\epsscale{0.5}
\epsfig{file=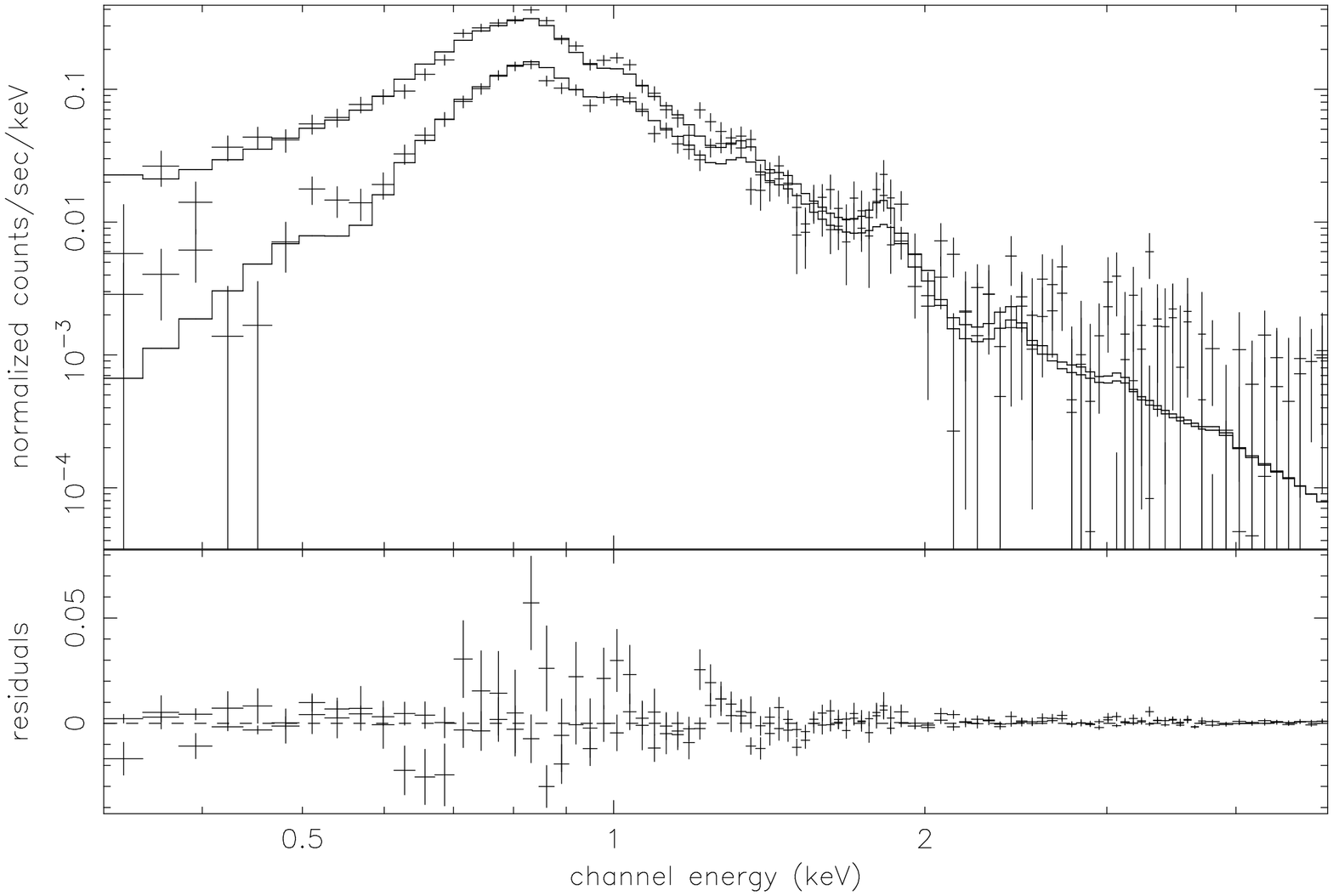,height=3in,width=3in,angle=270}
\hspace{0.3cm}\epsfig{file=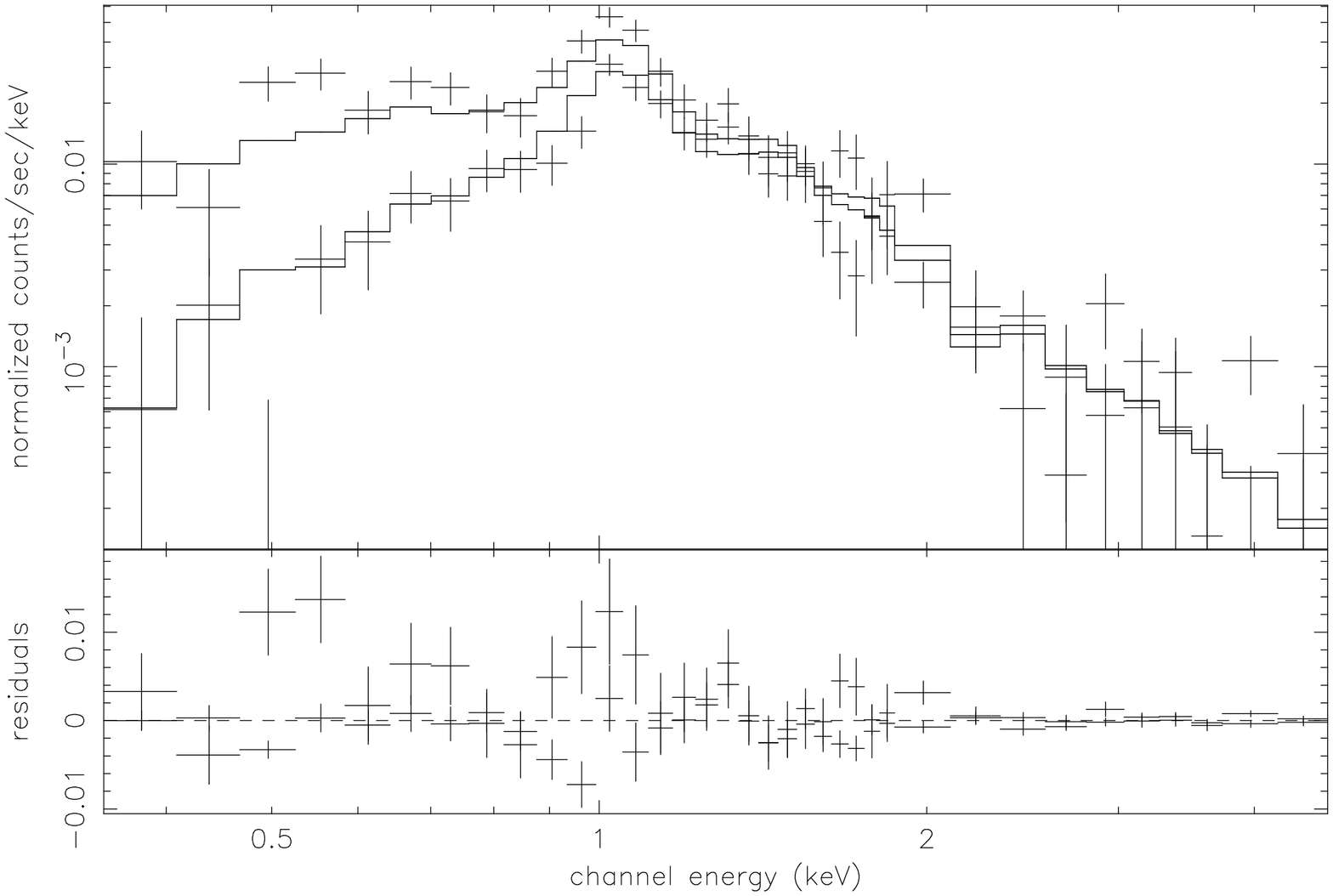,height=3in,width=3in,angle=270}
\caption{(left) Best simultaneous two temperature APEC model fit
to Obsid 2942 and Obsid 4174 data for the elliptical annulus (E  shown in 
Figure \protect\ref{fig:wedgegeom}) just inside the edge where 
the second component models the cluster foreground emission. 
 (right) Best simultaneous single temperature APEC model fit to 
 Obsid 2942 and Obsid 4174 data for the 
Fornax cluster gas in the free stream region (W$1$
shown in Figure \protect\ref{fig:wedgegeom}). Absorption is fixed at
the Galactic value ($n_{\rm H}=1.45 \times 10^{20}$\cms) in each region.
}
\label{fig:edgespec}
\end{figure}

\begin{deluxetable}{ccccccc}
\tablewidth{0pc}
\tablecaption{APEC Model Spectral Fits \label{tab:spectra}}
\tablehead{
\colhead{Region} & \colhead{R$_{\rm in}$,R$_{\rm out}$}  &
\colhead{Source II(III)} &\colhead{$kT$} &\colhead{$A$}  
  & \colhead{$\Lambda$}&\colhead{$\chi^2/{\rm dof}$} \\
 & kpc,kpc & counts & keV & $\Zs$ & $10^{-23}$erg\,cm$^{3}$s$^{-1}$ & }
\startdata
NGC~1404 E &$5.1$,$8.2$ & $4071(2790)$ &   &  &  &    \\
  $1$ component  &  &  &$0.58 \pm 0.01$  &$0.32 \pm 0.05$  
  &$1.0$  &$337/187$    \\
 $2$ component  &  &  &$1.53^{\rm f}$  &$0.42^{\rm f}$&  &     \\ 
                &  &  &$0.55^{+0.01}_{-0.02}$ &$0.73^{+0.65}_{-0.16}$
   & $1.9$ & $276/185$    \\ 
Fornax W$0$ &$8.2$,$13$  &$333(328)$  & $1.66^{+0.25}_{-0.15}$  
 &$0.42^{\rm f}$ &   & $11.2/20$   \\
Fornax W$1$ &$13$,$21$ &$930(710)$ &$1.53^{+0.10}_{-0.13}$
&$0.42^{+0.2}_{-0.13}$  
  &$0.846$ & $104.5/69$    \\
\enddata
\tablecomments{ Column $2$ lists the (inner,outer) radii of each extraction 
region shown in \protect\ref{fig:wedgegeom} measured from the center of 
the bounding ellipse. Source II (III) are the background
subtracted source counts in the $0.3-5$\,keV energy band 
for Obsid 2942 (Obsid 4174), respectively. Superscript $f$ denotes a fixed
parameter. $\Lambda$ denotes the X-ray emissivity (cooling function) 
derived from XSPEC for the $0.3-2$\,keV energy band.
Errors are $90\%$ confidence limits. 
 }
\end{deluxetable}
The results of the spectral analyses are summarized in Table 
\ref{tab:spectra}. The best fits for the cluster free-stream region
(W1) and for the galaxy emission inside the edge (E) are  shown in 
Figure \ref{fig:edgespec}. We find a temperature 
($1.53^{+0.10}_{-0.13}$\,keV) and abundance
($0.42^{+0.2}_{-0.13}\,\Zs$) for gas in 
the free stream region. These values  
are in excellent agreement with XMM-Newton measurements by 
Buote (2002) for diffuse gas $\sim 8'$ from NGC~1399 
($\sim 2'$ from NGC~1404), who finds the data at large radius
from NGC~1399 well represented by a single temperature APEC model with   
temperature $kT = 1.35 \pm 0.15$\,keV (errors $1\sigma$) and 
abundance $A \sim 0.4 - 0.6\,\Zs$. They are also in reasonable
agreement with 
Jones \etal (1997), who find 
$kT = 1.26^{+0.08}_{-0.11}$\,keV and $A=0.51^{+0.23}_{-0.16}\,\Zs$ for
radii $r \sim 8'-10'$ from NGC~1399 and by O'Sullivan \etal (2003), 
who find at the same radius 
$kT \sim 1.2-1.45$\,keV and $A \sim 0.4 - 0.7\,\Zs$ 
(see their Figure 1) using ROSAT PSPC data, and with Scharf \etal 
(2004) who find $kT \sim 1.6$ using Chandra ACIS-I data.

Due to limited statistics we are unable to freely vary all the
parameters of the fit for region W0, the diffuse gas just outside the
edge containing the stagnation region. We fix the abundance at the 
best fit value for region W1 and allow the temperature to vary. We
find a temperature $1.66^{+0.25}_{-0.15}$\,keV consistent with that
for region W1 given the large errors. The somewhat cooler temperature 
($\sim 1$\,keV) measured by O'Sullivan \etal (2003) with ROSAT data 
when using a circular annulus with mean radius $2'$ ($11$\,kpc) 
centered on NGC~1404 is most likely due to the fact that such a
circular region contains cooler ram-pressure stripped gas from 
the trailing side of NGC~1404 (the debris tail) as well as 
the Fornax ICM.

A single temperature APEC model
provides a poor fit to the data inside the edge, as shown in Table 
\ref{tab:spectra}. This is not surprising since a second component is
warranted to model the contribution from foreground cluster emission.
We model this emission with a separate APEC component with abundance 
and temperature fixed at the best fit cluster values. We find a
temperature for the galactic diffuse gas inside the edge of 
$0.55^{+0.01}_{-0.02}$\,keV with an abundance 
$0.73^{+0.65}_{-0.16}\,\Zs$. The fit is 
unchanged if the cluster component temperature is allowed to vary, 
although the errors on the fitted cluster temperature 
($1.48^{+0.24}_{-0.22}$\,keV) are large. The temperature we measure
for diffuse gas in NGC~1404 is in good agreement with previous 
measurements given the differences 
in the models and spectral extraction regions. Using ROSAT PSPC 
data, Jones \etal (1997) found a mean temperature for NGC~1404 of 
$0.65^{+0.02}_{-0.01}$\,keV; while O'Sullivan \etal (2003) found a 
temperature profile consistent with isothermal with $kT = 0.6 \pm
0.01$ for radii $r \lesssim 1'.6$, similar to that found by Scharf 
\etal (2004) using ACIS-I data.
We find, however, an 
abundance $A$ significantly higher than that found by most 
previous authors ($A \sim 0.14\,\Zs$, Loewenstein \etal 1994; 
$A \sim 0.16\,\Zs$, Jones \etal 1997; $A \sim 0.35$, O'Sullivan
\etal 2003), and more consistent with solar or super-solar 
abundances expected for elliptical galaxies (Buote 2002; 
Brighenti \& Mathews 1999). This discrepancy is probably due to 
a combination of limited statistics, large heterogeneous 
extraction regions, and the use of simple single
temperature models in those previous studies.  
We are in agreement with the abundance measurement 
($A = 0.7^{+2}_{-0.3}$) for NGC~1404 found by Buote \& Fabian (1998) 
using a multi-temperature MEKAL model to describe the ASCA data. 

In Table \ref{tab:spectra} we also list the X-ray emissivity 
(cooling function)  
$\Lambda$ derived from the XSPEC best fit models for the energy 
range ($0.3 - 2.0$\,keV), the energy band  we used for the 
surface brightness profiles in Section \ref{sec:edges}. 
In contrast to cluster 
analyses where the X-ray emissivity is dominated by the continuum
free-free emission  and depends only weakly on temperature and abundance 
(Vikhlinin \etal 2001), the 
emissivity of cool, metal-rich gas in galaxies is dominated by 
line emission and in this case varies by more than a factor of 
$2$ between X-ray emitting gas in the cluster and that in the 
galaxy. Since the X-ray surface brightness depends on the product of the 
emission measure and the X-ray emissivity, high quality abundance 
measurements in galaxies are crucial for determining accurate 
electron densities from surface brightness profiles and subsequently
constraining the dynamical motion of the system.

\subsection{Fitting the Surface Brightness Edge}
\label{sec:edges} 

\begin{figure}[t]
\epsscale{0.7}
\epsfig{file=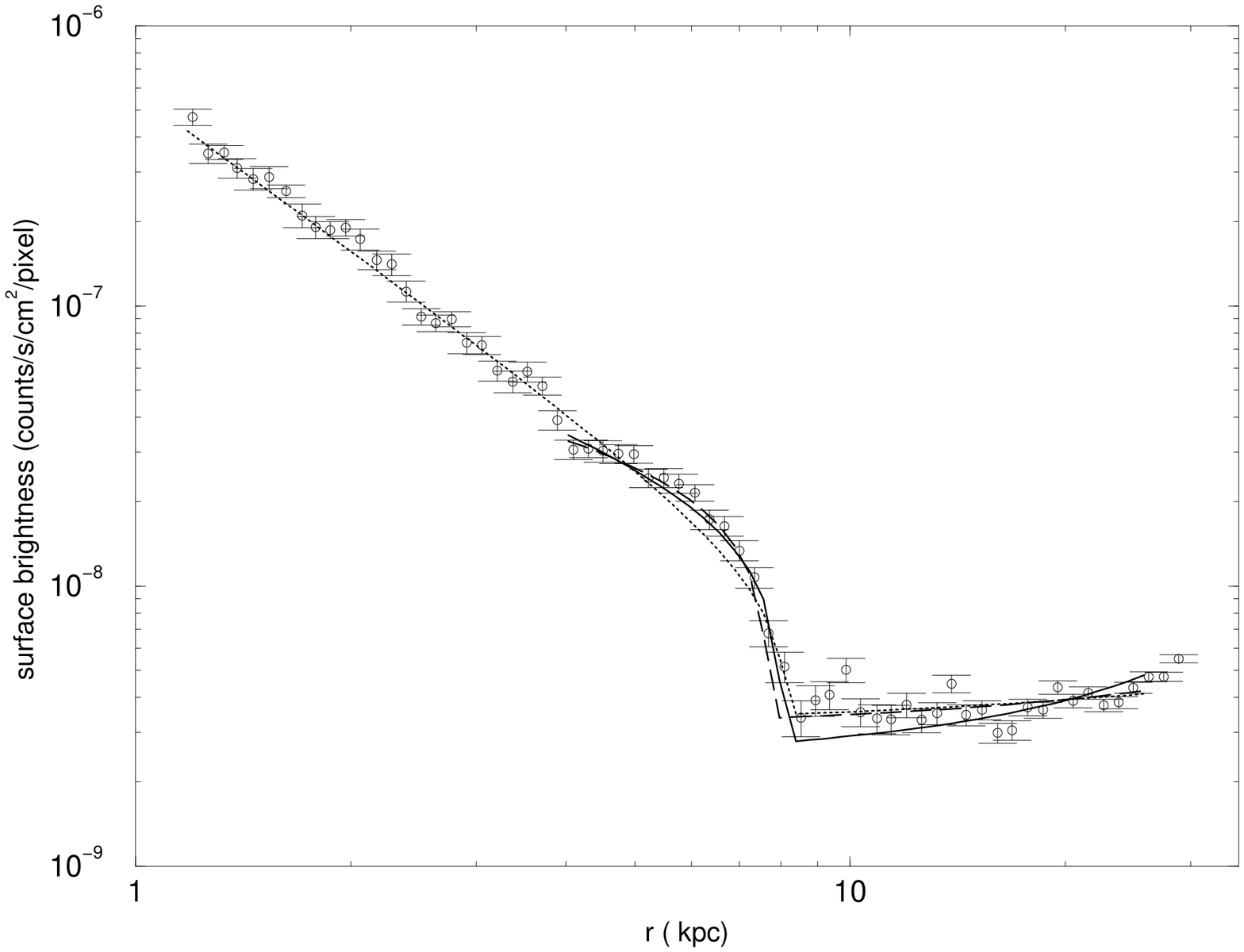,height=3in,width=3in,angle=270}
\hspace{0.3cm}\epsfig{file=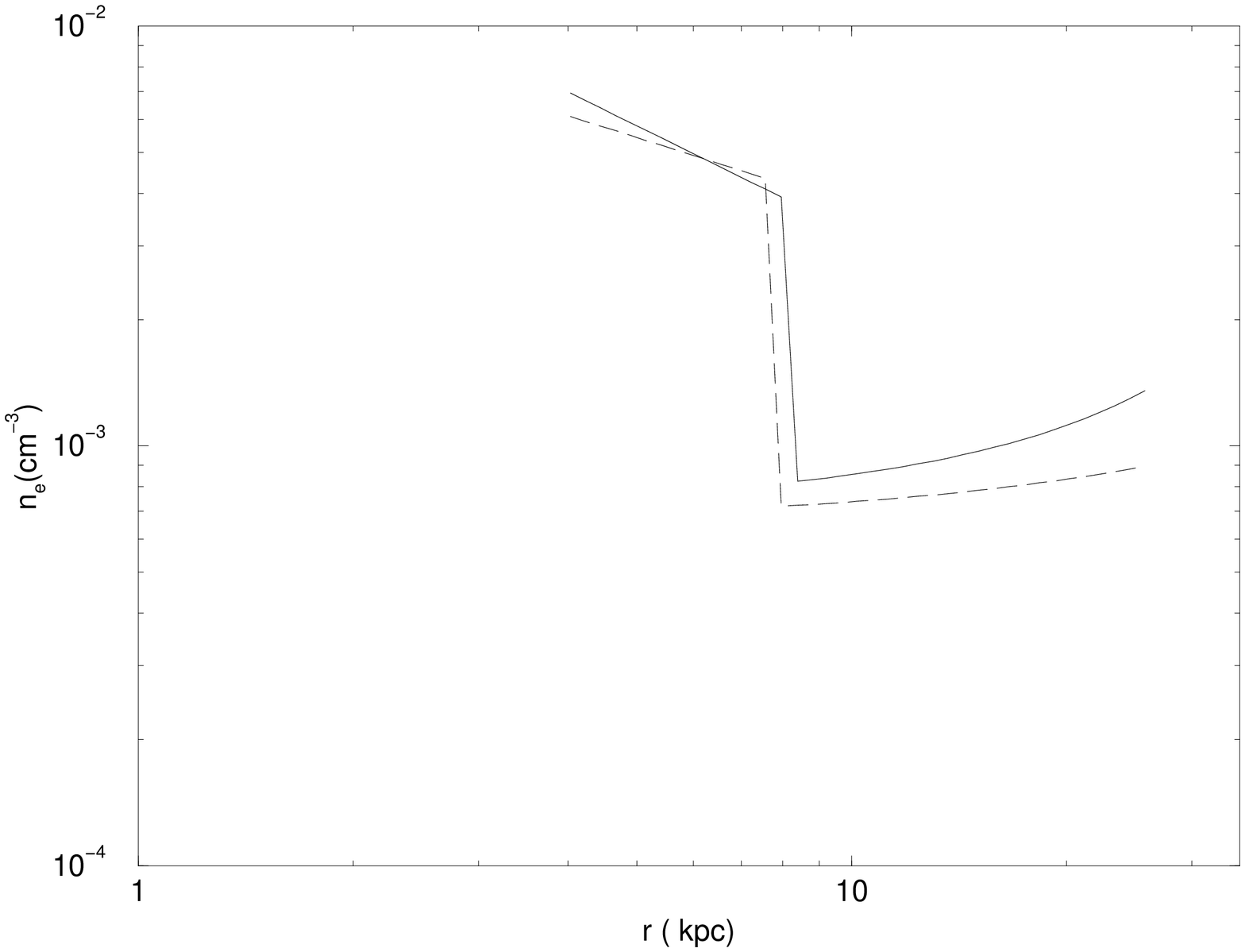,height=3in,width=3in,angle=270}
\caption{(left) Surface brightness profile across the edge from elliptical
annulli bounded by the sector shown in Figure
\protect\ref{fig:wedgegeom} as a function of the distance from the
center of the bounding ellipse. The models shown represent power law
density distributions inside the edge 
($4$\,kpc $\lesssim r \lesssim r_{\rm edge}$)
and two beta model fits to the Fornax ICM ($r > r_{\rm edge}$): 
$\beta=0.35$, $r_c=45''$ from Jones \etal (1995) (PL+ICM1, solid line)
and $\beta=0.3$, $r_c=394''$ (PL+ICM2, dashed line) from a three 
component beta model fit to NGC~1399. The dotted line shows the best 
fit for a single power law (SPL) model for radii 
$1$\,kpc $\lesssim r < r_{\rm edge}$ inside the edge and the ICM2 model
outside the edge.
(right) Corresponding electron density profiles for PL+ICM1 (solid line)
and PL+ICM2 (dashed line) models showing the electron density 
discontinuity across the edge.
}  
\label{fig:sbprof}
\end{figure}
\begin{figure}[t]
\epsscale{0.7}
\epsfig{file=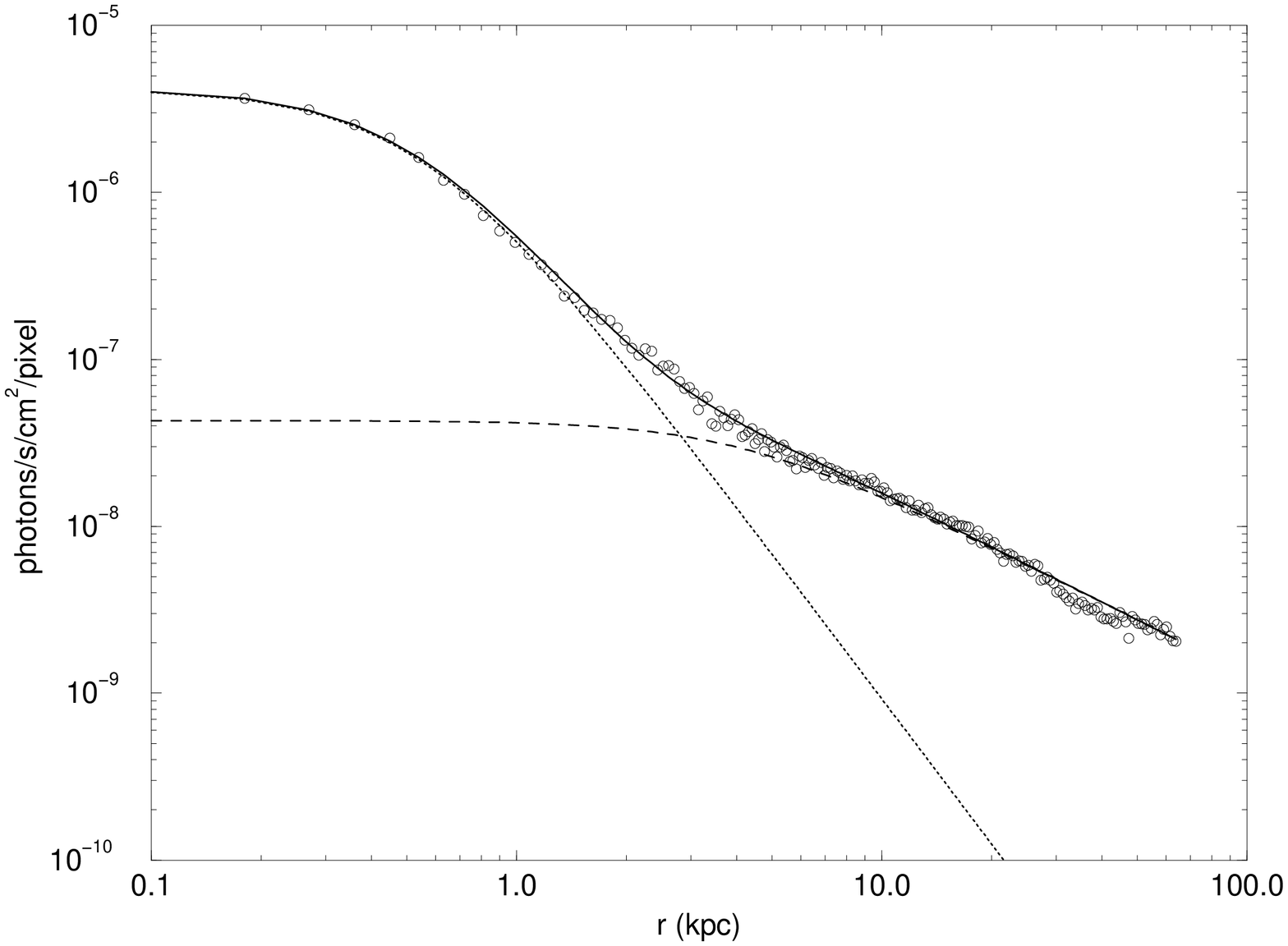,height=3in,width=3in,angle=270}
\hspace{0.3cm}\epsfig{file=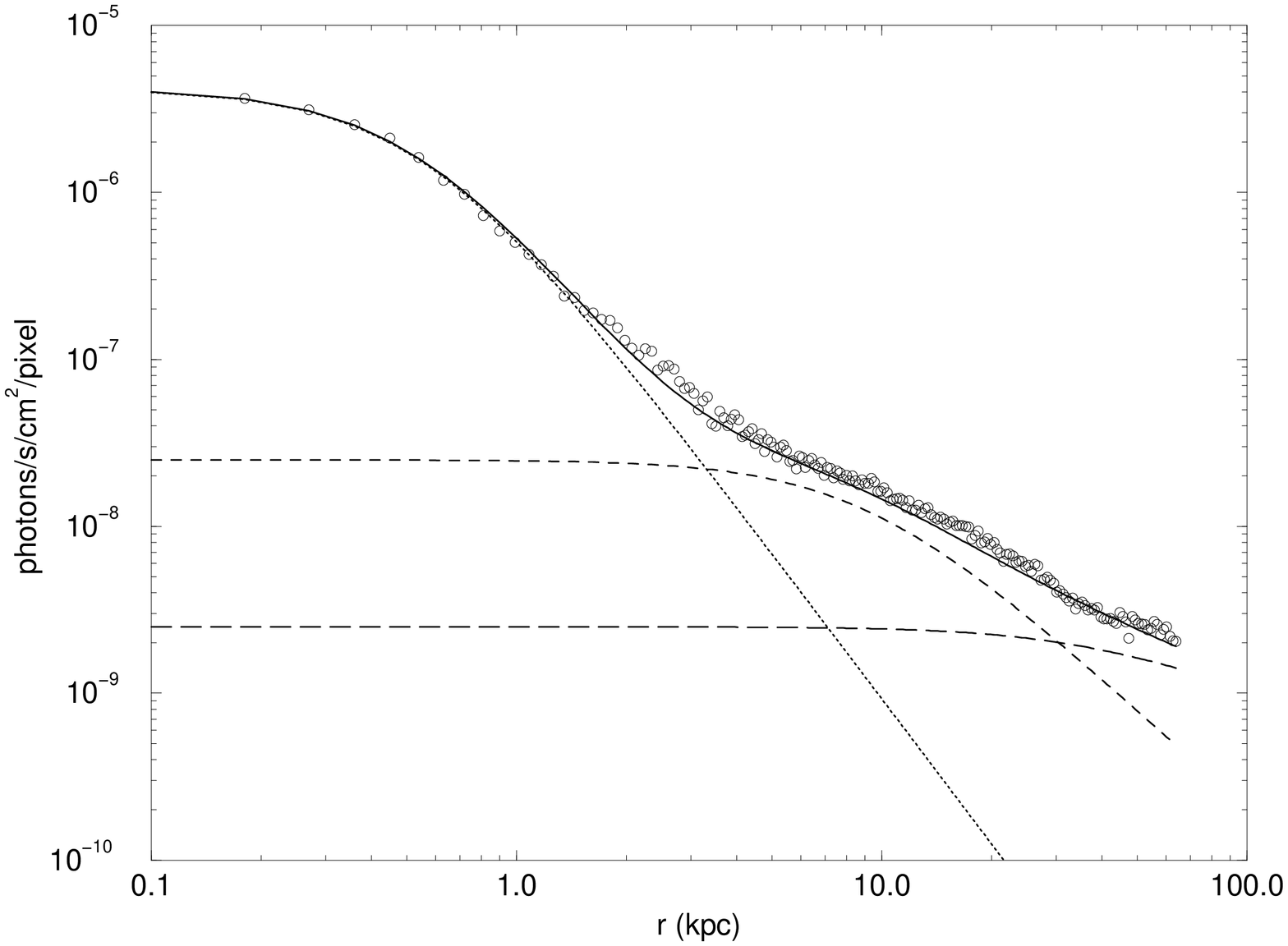,height=3in,width=3in,
angle=270}
\caption{Beta model fits to the surface brightness profile of
NGC~1399 as a function of the radial distance from the center of 
NGC~1399: (left) Two component fit with ($r_c,\beta$) given by  
($6'',0.65$, dotted line) and ($45'',0.35$, dashed line) for the core and 
outer regions, respectively.
(right) Three component fit with $r_c,\beta$ for the core, halo, and cluster
given by ($6'',0.65$, dotted line), ($98'',0.5$, dashed line), and 
($394'',0.3$,long dashed line), respectively. The solid line denotes the
sum of the components in each case.
}  
\label{fig:n1399rprof}
\end{figure}

In Figure \ref{fig:sbprof} we plot the X-ray surface brightness
profile for NGC~1404, constrained by the
wedge geometry shown in Figure \ref{fig:wedgegeom} to sample only the
region across the leading edge of the galaxy, as a function of the 
mean projected distance from the center of the edge-bounding ellipse.   
Three distinct regions are apparent. For small radii the data are well
described by a power law with a break at $\sim 4$\,kpc. For radii 
$4$\,kpc $\lesssim r \lesssim 8$\,kpc we see the surface brightness 
profile characteristic of a sharp edge, where the surface brightness 
decreases by $\sim$ an order of magnitude over a radial range of 
$\sim 1 - 2 $\,kpc. This is characteristic for motion of a spheroid 
through a uniform gas and is consistent with a jump-like density 
discontinuity at the boundary. Outside the edge ($r > 8$\,kpc) we see
a surface brightness profile consistent with cluster gas with a 
possible enhancement at $r \sim 10$\,kpc, near the edge. The gradual 
rise in surface brightness for $r > 20$\,kpc ($ \lesssim 34$\,kpc from
NGC~1399) is due to halo emission from NGC~1399 
(See Figure \ref{fig:n1399rprof}). 

In order to fit the surface brightness in Figure \ref{fig:sbprof}, we 
need to model the form of the electron density in the galaxy NGC~1404 
and the Fornax ICM. We assume, 
as in Vikhlinin \etal (2001), that the cold front inside the edge can 
be described by an isothermal sphere whose density is well modelled by
a power law 
\begin{equation}
n_e (r < r_{\rm edge})  
  = A_{\rm edge}\Bigl ( \frac{r}{r_{\rm edge}}\Bigr )^{-\alpha} 
\label{eq:densein}
\end{equation}
where $r_{\rm edge}$ is the position of the edge and 
$A_{\rm edge}$ is the electron density in the cold front inside the
galaxy at the position of the edge ($r=r_{\rm edge}$). 

The density model for the cluster emission is more problematical since
it requires knowledge of the density profile across the full extent of
the cluster along the line of sight ($r \lesssim 1.5$\,Mpc) and
may contain local variations due to the presence of significant
subcluster mass concentrations common in dynamically young clusters 
(see, e.g. Schindler \etal 1999 for
Virgo; Machacek \etal 2002 and AbdelSalam \etal 1998 for Abell 2218; 
Drinkwater \etal 2001 for Fornax). We adopt simple isothermal beta
models for the surface brightness and gas density of the Fornax
cluster gas outside the edge such that the 
surface brightness $S(r)$ is given by 
\begin{equation}
S(r)  = S_0\Bigl (1+\Bigl (\frac{r}{r_c}\Bigr )^2 \Bigr )^{-3\beta+0.5}.
\label{eq:betamodel}
\end{equation} 
where $S(r)$ is the surface brightness at a given radius, $S_0$ is the
central surface brightness, and $r_c$ is the core radius. Then the 
corresponding electron density is given by
\begin{equation}
n_e(r)  = 
   n_{0}\Bigl ( 1+\Bigl (\frac{r}{r_c}\Bigr )^2 \Bigr )^{-3 \beta/2} .
\label{eq:density}
\end{equation} 

In Figure \ref{fig:n1399rprof} we show two multiple-component beta
model fits to the surface brightness profile of NGC~1399, with point
sources and emission from NGC~1404 excluded, where the fit to the
data is the sum of individual components given by 
Equation \ref{eq:betamodel}. In each case the
innermost central region is modelled with $r_c = 6''$ and  
$\beta =0.65$. This result is in good agreement with the fit to
NGC~1399 by O'Sullivan \etal (2003), who find $r_c = 6.6 \pm 0.6''$ and 
$\beta=0.59 \pm 0.1$, although they use an elliptical beta model with 
axis ratio $1.23$ rather than our spherically symmetric one. 

Surface brightness profiles at larger radii have been measured by
Jones \etal (1995) and Paolillo \etal (2003) using 
ROSAT PSPC data. Jones \etal (1997) found the surface brightness
profile beyond the core of NGC~1399 well represented by a single beta model
with core radius $r_c = 45''$ and $\beta = 0.35$ out to a radius of 
$40'$($220$\,kpc). Paolillo \etal (2003) modelled the surface
brightness profile of NGC~1399 and the Fornax ICM with a three
component beta model, whose third (cluster) component dominated
emission beyond $7'$. Their model is consistent with the model of 
Jones \etal (1997) for radii $r \lesssim 200$\,kpc, where the ROSAT 
PSPC corrections are well known, but falls off more rapidly at larger
radii. 

Similarly,
we find our data equally well described by either a two or three
component beta model over the radial range of our observation. 
The left panel of Figure \ref{fig:n1399rprof} models the data with 
 a two component beta model where the outer component is 
parameterized by $r_c=45''$ and $\beta=0.35$, hereafter called 
ICM1, as found by Jones \etal (1997). The right panel
shows a three component beta model fit to the same data, where the
surface brightness at large radii is described by the sum of a central
component 
with ($r_c, \beta$) the same as above and two additional components, 
the galactic halo gas with  ($r_c=98''$, $\beta =0.5$) and cluster gas 
with  ($r_c=394''$, $\beta = 0.3$), hereafter called ICM2.   
The surface brightness edge in NGC~1404 is located at a 
projected distance of $\sim 46$\,kpc from the center of NGC~1399 where
the second (third) beta model component in the two (three) component
fit dominates. Since this represents the minimum physical distance 
between the edge and the center of the dominant cluster elliptical
NGC~1399 (and the overlap between components in this region is small),
we adopt the density beta model from Equation \ref{eq:density} 
with parameters fixed by those components as representative of the 
Fornax cluster gas in our analysis of the edge. Our data do  not
probe cluster distances large enough to distinguish between these
two density models. Thus we consider both models 
in order to investigate how uncertainties
in the cluster density distribution at large distances affect 
our results. 

In order to extract the density distribution from the surface
brightness profiles, we need to make additional assumptions about
the relative three dimensional geometry of NGC~1404 with respect 
to NGC~1399. The  
comet-like tail to the southeast of NGC~1404 seen in Figures 
\ref{fig:fornax} and \ref{fig:wedgegeom} is evidence for 
ram pressure stripping of gas from NGC~1404. Since 
ram pressure stripping is more likely in the denser, central regions
of the cluster (i.e. near the dominant elliptical
galaxy NGC~1399), we assume that NGC~1404 and NGC~1399 both lie near 
the plane of the sky such that the projected distance $9'.8$ is 
representative of their physical radial separation.  The direction of the 
tail, pointing to the southeast and away from the direction of motion,
indicates that NGC~1404 is moving to the northwest  
toward NGC~1399. Finally, the sharpness of the edge suggests
that we are able to directly view the stagnation region, 
such that the inclination of the motion with 
respect to the plane of the sky should not be too large (Mazzotta
\etal 2001). 
These are the same assumptions used in the analysis of the infalling 
subcluster in Abell 3667 (Vikhlinin \etal 2001).  Our analysis 
method closely follows that work. 

We calculate the surface brightness by integrating the 
square of the density times the corresponding X-ray emissivity, given 
in the sixth column of Table \ref{tab:spectra}, along the line of
sight. The data 
are fit using a multi-variate chi-square minimization scheme allowing
the position of the edge ($r_{\rm edge}$), density power law index inside
the edge ($\alpha$), and galaxy electron density at the edge 
($A_{\rm edge}$) to vary.  We also allow the cluster density 
normalization ($n_{0}$ in Equation \ref{eq:density}) to be a free 
parameter to partially compensate for possible local density
fluctuations in the cluster ICM when extracting the cluster 
electron density near the edge. 

In Figure \ref{fig:sbprof} we show the results for three density 
models: a power law (PL) density model for the edge region 
($4$~kpc~$< r < r_{\rm edge}$) interior to NGC~1404 with the cluster 
density modelled by either ICM1 or ICM2, and a single power law 
(SPL) model for $r > 1$\,kpc, with no break at $r \sim 4$\,kpc. 
The ICM2 model is used for the cluster density in this latter case.
 As the dotted line in 
Figure \ref{fig:sbprof} shows, the surface brightness profile for
radii $\lesssim 4$\,kpc is well reproduced by this simple power law 
density model for the galaxy  (SPL+ICM2) with a best fit power law 
 index $\alpha = 1.54$ (corresponding to a beta model with $\beta = 0.47$
for $r >> r_c$), in reasonable agreement with the beta model results
($\beta = 0.51$, $r_c = 0.57$\,kpc) for NGC~1404 by Paolillo \etal
(2002) and ($\beta \sim 0.5$) by Scharf \etal (2004).
However, this model is unable to reproduce the shape of the surface 
brightness distribution in the edge region 
($4$~kpc~$< r < r_{\rm edge}$). 
Both density models PL+ICM1 (solid line) and PL+ICM2 (dashed line) 
reproduce the surface brightness profile across the edge well. The fit
parameters for these models are summarized in Table \ref{tab:fits}. 
The electron density we derive by fitting the surface brightness
profile directly for gas inside the edge ($3.9 \times 10^{-3}$\cmc for
the PL+ICM1 model and $4.3 \times 10^{-3}$\cmc for the PL+ICM2 model) 
are in excellent agreement with the beta model result 
($\sim 4 \times 10^{-3}$\cmc) result by Scharf \etal (2004), although
our value for the cluster gas is $\sim 20\%$ higher.  

\begin{deluxetable}{ccc}
\tablewidth{0pc}
\tablecaption{Best Fit Edge Position and Electron Densities\label{tab:fits}}
\tablehead{
\colhead{Property} & \colhead{PL+ICM1 Model}  &
\colhead{PL+ICM2 Model} }
\startdata
$r_{\rm edge}$ (kpc)& $8.01$ & $7.64$ \\
$\alpha$ & 0.84  & 0.54  \\
$n_{\rm in}$ ($10^{-3}$\cmc) & 3.9  & 4.3  \\
$n_{\rm out}$ ($10^{-3}$\cmc) & 0.82  & 0.72 \\
\enddata
\tablecomments{ $r_{\rm edge}$ is the position of the edge measured
from the center of the bounding ellipse in 
Figure \protect\ref{fig:wedgegeom}, $\alpha$ is the power law index
of the spherically symmetric electron density distribution inside the 
edge, and $n_{\rm in}$ ($n_{\rm out}$) is the electron density just 
inside (outside) the edge. 
 }
\end{deluxetable}

\subsection{Bounding the Galaxy's Dynamical Motion}
\label{sec:velocity}

\begin{figure}[t]
\begin{center}
\epsscale{0.7}
\epsfig{file=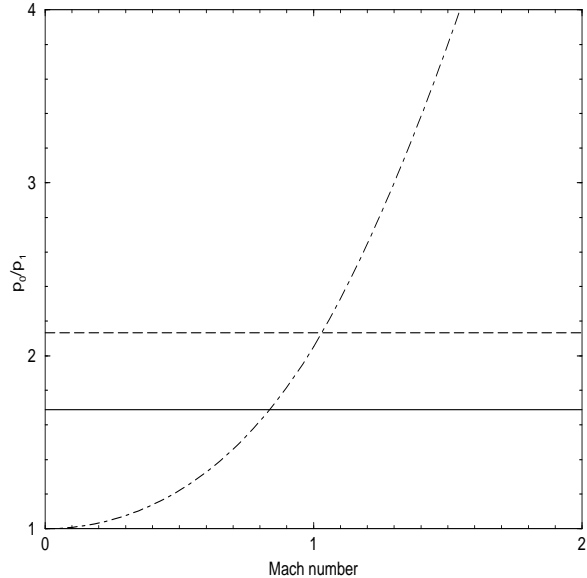,height=3in,width=3in,angle=270}
\caption{Pressure jump as a function of Mach number as given by 
 Equation \protect\ref{eq:subsonic} for $M_1 < 1$ and Equation 
\protect\ref{eq:mach} for $M_1 > 1$ (dot-dashed line). The horizontal 
lines denote the value of the pressure jump for the PL+ICM1 model
(solid line) and for the PL+ ICM2 model (dashed line) fits to the 
surface brightness profile across the edge.
}  
\label{fig:mach}
\end{center}
\end{figure}

Following Vikhlinin \etal (2001) we use
the densities and the temperatures from our surface 
brightness and spectral fits to calculate the gas pressures 
in the free stream region and in the galaxy gas just inside the edge, 
which is in pressure equilibrium with gas at the stagnation
point (where the relative cluster gas velocity vanishes). 
The difference between the two pressures is interpreted as
the ram pressure of the intracluster medium on the leading edge of the
galaxy (the cold front); while the 
ratio of the pressures allows us to determine the Mach number
$M_1=v/c_1$ (where $c_1$ is the speed of sound in the cluster free
stream region) for the cold gas cloud moving through the hot ICM. For 
completeness, the pressure-jump versus Mach number relation, plotted
in Figure \ref{fig:mach}, is given
by :
\begin{equation}
\frac{p_0}{p_1} = \Bigl (1 + \frac{(\gamma -1)}{2}M_1^2 \Bigr )^{(\frac{\gamma}
{\gamma -1})}
\label{eq:subsonic}
\end{equation}
for $M_1 \leq 1$ (subsonic regime); while
\begin{equation}
\frac{p_0}{p_1} = \Bigl (\frac{\gamma +1)}{2} \Bigr )^{(\gamma +
1)/(\gamma -1)}M_1^2 \Bigl [ \gamma - \frac{\gamma -1}{2 M_1^2}\Bigr ]
\label{eq:mach}
\end{equation}
for $M_1 > 1$ (supersonic regime), with 
$\gamma = 5/3$ the adiabatic index for a monatomic, ideal gas 
(Landau \& Lifshitz, 1959). 

\begin{deluxetable}{ccc}
\tablewidth{0pc}
\tablecaption{Derived Velocities and Edge Properties for Two Cluster
ICM Models\label{tab:properties}}
\tablehead{
\colhead{Property} & \colhead{PL+ICM1 Model}  &
\colhead{PL+ICM2 Model} }
\startdata
$n_{e0}/n_{e1}$ & $4.78$ & $6.04$ \\
$T_0/T_1$ & $0.36$ & $0.36$ \\
$p_0/p_1$ & $1.7$ & $2.1$ \\
$M_1$ & $ 0.83$ & $1.03$ \\
$v$(kms$^{-1}$) & $531$ & $657$ \\
$v_{\rm t}$(kms$^{-1}$) & $98$($298$) & $399$($478$) \\
$\xi$ (degrees) & $79$($58$) & $53$($43$) \\
\enddata
\tablecomments{$n_e$, $T$, and $p$ denote the electron density, temperature
and pressure in each region specified by subscripts $0$ and $1$ 
for galaxy gas just inside the edge and cluster gas in the free 
stream region, 
 respectively (See Vikhlinin \etal 2001). $M_1$ is the Mach
number, $v$ is the relative total velocity, and $v_{\rm t}$ and
$\xi$ are the component of velocity in the plane of the
sky and inclination angle of motion with respect to the plane of the
sky, respectively, assuming a relative radial velocity between
NGC~1404 and NGC~1399 of $v_r = 522$\kms
from NED ($v_r = 450$\kms from Drinkwater \etal 2001).
 }
\end{deluxetable}
In Table \ref{tab:properties} we summarize the results of this analysis.
From Figure \ref{fig:mach} we find a Mach number for the galaxy gas
cloud moving through the
cluster gas of $0.83$ ($1.03$) for the PL+ICM1 (PL+ICM2)
best fit density models, respectively. The sound speed for the
completely ionized (mean molecular weight $\mu=0.6$) intracluster gas
at a temperature $1.53^{+0.10}_{-0.13}$\,keV in the free stream region
is $c_1=638$\kms. Thus the speed $v=M_1c_1$ of NGC~1404  relative to the
ICM is found to be $v =531 (657)$\,kms for these two models, in
excellent agreement with the estimate ($v \sim 660 \pm 260$\kms) by 
Scharf \etal (2004).
If we assume that the Fornax cluster ICM, at least that part of it 
that impacts the leading edge of NGC~1404, is at rest relative to
NGC~1399 and use the radial (line of sight) velocities $1947$\kms for 
NGC~1404 and $1425$\kms for NGC~1399 given in NED (Graham \etal 1998),
then the radial velocity between these galaxies would be $522$\kms. The
relative transverse component of the velocity between NGC~1404 and 
the Fornax ICM is then $98$ ($399$)\kms, respectively, implying an 
inclination angle for the motion with respect to the plane of the sky
of $\xi \sim 79^\circ$ ($53^\circ$), respectively, for our two models.
However, Mazzotta \etal (2001) stress that the sharpness of the edge 
is a strong function of the inclination angle $\xi$. Given 
the sharpness of the observed  edge in NGC~1404, these derived
inclination angles are uncomfortably large. If we adopt instead
the optical relative radial velocity 
$v_r = 450$\kms measured by Drinkwater \etal (2001), who identify 
NGC~1404 as part of a larger, high velocity, infalling Fornax 
subcluster, the inferred transverse velocities for NGC~1404 
increase to $\sim 292$ ($479$)\kms, respectively, for our two 
models.  While improved,   
the inferred inclination angle of the motion with respect to the 
plane of the sky, $\xi = 58^\circ$ ($43^\circ$), is still large. 

There are several possible solutions to this dilemma. First, NGC~1404 
and NGC~1399 may not both lie in the plane of the sky. NGC~1404 may be
undergoing a fly-by of the dominant elliptical, with a non-negligible 
impact parameter  rather than a direct
collision, such that the use of 
the projected distance between the galaxies underestimates the 
physical radial distance between them and overestimates the 
value of the cluster electron density at the edge. Thus the density jump 
should be viewed as a lower bound on the physical density
discontinuity, Mach number, and relative velocity of NGC~1404 
with respect to the intracluster gas. A second possibility is that 
NGC~1399 is moving with respect to the Fornax ICM and may, itself, 
be experiencing ram pressure stripping due to this motion 
(Paolillo \etal 2002). Then the optically measured relative
radial velocity between the two galaxies may not accurately reflect the
relative radial velocity between NGC~1404 and the surrounding cluster
gas. 

Despite these uncertainties it is useful to consider what our  
analysis implies for the increased temperature and density expected
in the stagnation region. For the subsonic motion ($M = 0.83$) 
found for model PL+ICM1 and using the cluster free stream 
as the reference region, Bernoulli's equation predicts a density 
increase of $37\%$ leading to a surface brightness enhancement by a 
factor of $\sim 1.9$. This is
consistent with the possible $\sim 70\%$ surface brightness enhancement
seen justoutside the edge in Figure \ref{fig:sbprof}. The temperature in the
stagnation region would be expected to rise by $\sim 23\%$ by
adiabatic compression to $\sim 1.88$\,keV. However, the stagnation 
region is small (of the order of a galaxy diameter) such that the observed 
temperature would be lower because of dilution by cooler cluster 
gas along the line of sight. Our measurements for the temperature
in this region hint of a possible temperature rise consistent with
such adiabatic heating, but our errors are too large because of  
limited statistics for a definitive measurement. For model PL+ICM2 
($M \sim 1.03$) the temperature and density behind the shock would 
be expected to increase by $\sim 3\%$ and $\sim 5\%$, respectively, 
with temperature and density in the stagnation region similar to 
that for model PL+ICM1. 
Such a weak shock would not be expected to be seen in our 
data. 

\section{Conclusions}
\label{sec:conclude}

In this paper we have used data from three Chandra observations using
the ACIS array (a $57.4$~ksec exposure with chips S2 and S3
taken on 18-19 January 2000, a $29.6$\,ks exposure using
chip S3 taken on 13 February 2003 , and a $47.3$\,ks exposure 
using the ACIS-I array taken on 28-29 May 2003)  of 
the elliptical galaxy NGC~1404 falling toward the dominant elliptical
galaxy NGC~1399 in the Fornax cluster to bound the relative physical 
motion of NGC~1404 with respect to the Fornax cluster ICM. We find: 
\begin{enumerate}
\item{The temperature and abundance of the galaxy gas just inside the
edge in NGC~1404 is $0.55^{+0.01}_{-0.02}$\,keV and 
$0.73^{+0.65}_{-0.16}\Zs$, typical of elliptical galaxies.
Gas temperatures for the cluster gas in the free-stream region 
(W1) and region next to the edge containing the stagnation point (W0)
are
$1.53^{+0.10}_{-0.13}$\,keV and 
$1.66^{+0.25}_{-0.15}$\,keV, respectively, 
with a cluster abundance $0.42^{+0.2}_{-0.13}\Zs$ measured in
W1.} 

\item{The X-ray emissivity in NGC~1404 is more than a factor
$2$ greater than that of the surrounding cluster gas. Since the
X-ray emissivity is dominated by line emission in cool, metal rich 
galaxies, high quality abundance measurements in galaxies are 
crucial for determining the density discontinuity across the edge
and constraints on the dynamical motion of the galaxy through
the ICM from these edge analyses.}

\item{The surface brightness profile in NGC~1404 in the direction towards
NGC~1399 obeys a power law for $1\,{\rm kpc} < r < 4\,{\rm kpc}$, 
a break at $\sim 4$\,kpc, and a sharp edge where the surface
brightness decreases by more than an order of magnitude over 
$1 - 2$\,kpc. Outside the edge the surface brightness profile is 
consistent with cluster gas with a possible $50 - 70\%$ 
enhancement at $r \sim 10$\,kpc.}

\item{The best fit edge position is
found to be at 
$7.6 - 8.0$\,kpc from the center of the edge-bounding
ellipse with a weak dependence on the 
large distance behavior of the cluster gas density. A single power 
law density model within NGC~1404 provides a 
poor fit to the data for $r \gtrsim 4$\,kpc. The surface 
brightness profile across the edge is well modelled
with a power law electron density in NGC~1404 with index 
$\alpha = 1.54$ (beta 
model index $\beta = 0.47$) for $1\,{\rm kpc} < r < 4\,{\rm kpc}$, 
a break at $\sim 4$\,kpc after which $\alpha$ flattens to 
$ \sim 0.5 - 0.8$ with a weak dependence on the beta model parameters
assumed for the cluster gas. The sharpness of the edge is consistent
with a jump-like density discontinuity with electron density in NGC~1404 
inside the edge of $3.9 - 4.3 \times 10^{-3}$\cmc and in the cluster 
free stream region of $7 - 8  \times 10^{-4}$\cmc. }

\item{The pressure jump, $p_0/p_1 \sim 1.7 - 2.1$, 
between the cluster free stream region and
gas at the edge implies a Mach number 
$M_1 \sim 0.83 - 1.03$ relative to gas in the cluster free stream region
and thus a velocity $v \sim 531 - 657$\kms for NGC~1404 through the 
surrounding ICM. This motion predicts an enhancement 
in the surface brightness of the cluster gas in the stagnation region 
by a factor $\sim 1.9$ due to adiabatic compression,  
consistent with that seen. }

\item{Using values for the relative radial velocity between NGC~1404
and NGC~1399 drawn from the literature as representative of the 
relative radial velocity between NGC~1404 and the surrounding cluster 
gas, the inferred value for the inclination angle of the motion 
with respect to the plane of the sky is large, $\gtrsim
40^\circ$, in
apparent contradiction with the observed sharpness of the edge. This
may signal the presence of a significant impact parameter between 
NGC~1404 and NGC~1399 or that NGC~1399 is also moving radially 
with respect to the cluster gas. Thus, more detailed  hydrodynamic 
modelling of this system is required to resolve the full 
three dimensional motion of this system.}

\end{enumerate}


\acknowledgements

This work is supported in part by NASA contract NAS8-03060 
and the Smithsonian Institution. MEM acknowledges partial support from
the Radcliffe Institute for Advanced Study at Harvard University. 
This work has made use of the NASA/IPAC Extragalactic Database (NED)
which is operated by the Jet Propulsion Laboratory, California
Institute of Technology,  under contract with the National
Aeronautics and Space Administration. We wish to thank Paul Nulsen
for helpful discussions.

\begin{small}

\end{small}
\vfill
\eject
\end{document}